\documentclass[letterpaper, 10 pt, conference]{ieeeconf}  

\IEEEoverridecommandlockouts                              
\usepackage{graphicx}
\usepackage{amsmath}

\newtheorem{theorem}{\bf Theorem}
\newtheorem{corollary}{\bf Corollary}
\newtheorem{lemma}{\bf Lemma}
\newtheorem{definition}{\bf Definition}

\newtheorem{claim}{\bf Claim}

\newtheorem{example}{\bf Example}

\title{\LARGE \bf
Congestion Games with Resource Reuse and Applications in Spectrum Sharing
}
\author{Mingyan Liu, Sahand Haji Ali Ahmad and Yunnan Wu
\thanks{This work is supported by NSF award CNS-0238035 and through collaborative participation in the Communications and Networks Consortium sponsored by the U. S. Army Research Laboratory under the Collaborative Technology Alliance Program Cooperative Agreement DAAD19-01-2-0011.}
\thanks{M. Liu and S. Ahmad are with the Electrical Engineering and Computer Science Department, University of Michigan, Ann Arbor, MI 48105, USA
        {\tt\small \{shajiali,mingyan\}@autsubmit.com}}%
\thanks{Y. Wu is with Microsoft Research, Redmond, WA 98052, USA
        {\tt\small yunnanwu@microsoft.com}}%
}

\begin{document}

\maketitle
\thispagestyle{empty}
\pagestyle{empty}

\begin{abstract}

In this paper we consider an extension to the classical definition of congestion games (CG) in which multiple users share the same set of resources and their payoff for using any resource is a function of the total number of users sharing it.  The classical congestion games enjoy some very appealing properties, including the existence of a Nash equilibrium and that every improvement path is finite and leads to such a NE (also called the finite improvement property or FIP), which is also a local optimum to a potential function.  On the other hand, this class of games does not model well the congestion or resource sharing in a wireless context, a prominent feature of which is spatial reuse.  What this translates to in the context of a congestion game is that a user's payoff for using a resource (interpreted as a channel) is a function of the its number of its {\em interfering} users sharing that channel, rather than the total number among all users.  This makes the problem quite different.  We will call this the congestion game with resource reuse (CG-RR).  In this paper we study intrinsic properties of such a game; in particular, we seek to address under what conditions on the underlying network this game possesses the FIP or NE.  We also discuss the implications of these results when applied to wireless spectrum sharing.

\end{abstract}


\section{Introduction}

In this paper we present a generalized form of the class of non-coopertive strategic games known as {\em congestion games} (CG) \cite{Rosenthal1973,vocking2006cgo}, and study its properties as well as its application to wireless spectrum sharing.

A congestion game $\Gamma$ is given by the tuple $({\cal N}, {\cal R}, (\Sigma_i)_{i\in\cal{N}}, (g_r)_{r\in\cal{R}})$, where ${\cal N}=\{1, 2, \cdots, N\}$ denotes a set of players/users, ${\cal R}=\{1, 2, \cdots, R\}$ the set of resources, $\Sigma_i\subset 2^{\cal{R}}$ the strategy space of player $i$, and $g_r: {\mathsf N}\rightarrow{\mathsf Z}$ a payoff (or cost) function associated with resource $r$. Specifically $g_r$ is a function of the total number of users of resource $r$.  A player in this game aims to maximize (minimize) its total payoff (cost) which is the sum over all resources its strategy involves.  More detailed and formal description of this class of games are provided in Section \ref{sec-review}.

The congestion game framework is well suited to model resource competition
where the resulting payoff (cost) is a function of the level of congestion (number of active users).
It has been extensively studied within the context of network routing, see for instance the network congestion game studied in \cite{Fabrikant04}, where source nodes seek minimum delay path to a destination and the delay of a link depends on the number of flows going through that link.

Congestions games are closely related to potential games \cite{Monderer1996}, and enjoy some remarkable features.
 In particular, a congestion game is an exact potential game as it admits an exact potential function \cite{vocking2006cgo}.
Finding a solution (Nash equilibrium or NE) to a congestion game is equivalent to finding a (local) optimal solution to this potential function.  It is also known that any improvement path is finite (in which each player's improvement move also improves the potential)
and leads to a pure strategy NE.  In other words, while the system is decentralized and all players are selfish, by seeking to optimize their individual objectives they end up optimizing a global objective, the potential function, and do so in a finite number of step regardless of the updating sequence.\footnote{This in turn means that if the potential function of a particular congestion game has a meaningul and desirable physical interpretation, then the solution (an NE) to this decentralized game has certain built-in performance guarantee, as it is also a local optimal solution to a global objective.  This is a desirable feature as in general an NE can be fairly inefficient with respect to a given global objective function.}

With its appealing physical interpretation and the aforementioned attractive features, it is tempting to model resource competition in a wireless communication system as a congestion game.
However, the standard congestion game fails to capture two critical aspects of resource sharing in wireless communication: {\em interference} and {\em spatial reuse}.
A key assumption underlying the congestion game model is that all users have an equal impact on the congestion, and therefore all that matters is the total number of users of a resource\footnote{This function may be user-specific (see for example the one studied in \cite{milchtaich1996cgp}), but it remains a function of the total number of active users of that resource.}.
This however is not true in wireless communication.
Specifically, if we consider bandwidth or channels as resources, then sharing the same channel is complicated by pair-wise interference; a user's payoff (e.g., channel quality, achievable rates, etc.) depends on {\em who} the other users are and how much interference it receives from them.  If all other simultaneous users are located sufficiently far away, then sharing may not cause any performance degradation, a feature commonly known as spatial reuse.
%

The above consideration poses significant challenge in using the congestion game model depending on what type of user objectives we are interested in.  In our recent work \cite{liu-allerton08}, we tried to address the user-specific interference issue within the congestion game framework, by introducing a  concept called {\em resource expansion}, where we define virtual resources as certain spectral-spatial unit that allows us to capture pair-wise interference.  This approach was shown to be quite effective for user objectives like interference minimization.  In particular, using resource expansion we were able to demonstrate that two recently published distributed interference minimization algorithms in a multi-channel multi-user system \cite{Tarokh,Misra} have equivalent congestion game form representations, thereby showing that (1) stability and optimality results can be obtained automatically following this mapping, and (2) these problems can be made a lot more general by drawing from known results on congestion games.   The same idea also allows us to formulate a base station channel adaption problem in \cite{liu-allerton08}.


In this paper, we take a different approach where we generalize the standard congestion games to directly account for the interference relationship and spatial reuse in wireless networks.  Specifically, under this generalization, 
%
each user is associated with an interference neighborhood, and in using a resource (a wireless channel in our context), its payoff is a function of the total number of users {\em within its interference neighborhood} using it.  In other words, resources are {\em reusable} beyond a user's interference set in that the user is oblivious to users outside this set even if they are simultaneously using the same resources.  This extension is a generalization of the original congestion game definition, as the former reduces to the latter if all users in the game belong to exactly the same interference domain/neighborhood (i.e., every user interferes with every other user).  This class of generalized games will be referred to as {\em congestion games with resource reuse} (CG-RR).

The applicability of this class of games to a multi-channel, multi-user wireless communication system can be easily understood.  Specifically, we consider such a system where a user can only access one channel at a time, but can switch between channels.  A user's principal interest lies in optimizing its own performance objective (i.e., its data rate) by selecting the best channel for itself.  This and similar problems have recently captured increasing interest from the research community, particularly in the context of cognitive radio networks (CRN) and software defined ratio (SDR) technologies, whereby devices are expected to have far greater flexibility in sensing channel availability/condition and moving operating frequencies.

While directly motivated by resource sharing in a multi-channel, multi-user wireless communication system, the definition of CG-RR is potentially more broadly applicable.  It simply reflects the notion that in some application scenarios resources may be shared without conflict of interest.   In subsequent sections we will examine what properties this class of games possesses (in particular, under what conditions the finite improvement properties or a Nash equilibrium exists).

It has to be mentioned that game theoretic approaches have often been used to devise effective decentralized solutions to a multi-agent system.   Within the context of wireless communication networks and interference modeling, different classes of games have been studied. An example is the well-known {\em Gaussian interference game} \cite{Tse2007,Cioffi2002}.  In a Gaussian interference game, a player can spread a fixed amount of power arbitrarily across a continuous bandwidth, and tries to maximize its total rate in a Gaussian interference channel over all possible power allocation strategies.  It has been shown \cite{Tse2007} that it has a pure strategy NE, but the NE can be quite inefficient; playing a repeated game can improve the performance.  In addition, previous work \cite{Huang2006} investigated a market based power control mechanism via supermodularity, while previous work \cite{Goldsmith} studied the Bayesian form of the Gaussian interference game in the case of incomplete information.

By contrast, in our problem the total power of a user is not divisible, and it can only use it in one channel at a time.  This set up is more appropriate for scenarios where the channels have been pre-defined, and the users do not have the ability to access multiple channels simultaneously (which is the case with many existing devices).   In addition, in a CG-RR interference is modeled using the notion of interference set (equivalent of a binary interference relationship) whereas a Gaussian interference game interference is calculated using pair-wise distance.   These differences lead to very different technical approaches and results.

The organization of the remainder of this paper is as follows.  In Section \ref{sec-review} we present a brief view on the literature of congestion games, and formally define the class of congestion games with resource reuse in Section \ref{sec-problem}.  We then derive conditions under which this class of games possesses the finite improvement property (Section \ref{sec-fip}).  We further show a series of conditions, on the underlying network graph in Section \ref{sec-graph} and on the user payoff function in Section \ref{sec-payoff}, under which these games have an NE.  We discuss extensions to our work and conclude the paper in \ref{sec-conclusion}.

\section{A Review of Congestion Games}
\label{sec-review}

In this section we provide a brief review on the definition of congestion games, their relation to potential games and their known properties\footnote{This review along with some of our notations are primarily based on references \cite{Rosenthal1973,vocking2006cgo,Monderer1996}.}.
We then discuss why the standard congestion game does not take into account interference and spatial reuse, and motivate our generalized CG-RR games.

\subsection{Congestion Games}

Congestion games \cite{Rosenthal1973,vocking2006cgo} are a class of strategic games given by the tuple $({\cal N}, {\cal R}, (\Sigma_i)_{i\in{\cal N}}, (g_r)_{r\in{\cal{R}}})$, where ${\cal N}=\{1, 2, \cdots, N\}$ denotes a set of users, ${\cal R}=\{1, 2, \cdots, R\}$ a set of resources, $\Sigma_i\subset 2^{\cal{R}}$ the strategy space of player $i$, and $g_r: {\bf N}\rightarrow{\bf Z}$ a payoff (or cost) function associated with resource $r$.
The payoff (cost) $g_r$ is a function of the total number of users using resource $r$ and in general assumed to be non-increasing (non-decreasing).  A player in this game aims to maximize (minimize) its total payoff (cost) which is the sum total of payoff (cost) over all resources its strategy involves.

If we denote by $\sigma=(\sigma_1, \sigma_2, \cdots, \sigma_N)$ the strategy profile, where $\sigma_i\in \Sigma_i$, then user $i$'s total payoff (cost) is given by
\begin{eqnarray}
g^i(\sigma) = \sum_{r\in \sigma_i} g_r(n_r(\sigma))
\end{eqnarray}
where $n_r(\sigma)$ is the total number of users using resource $r$ under the strategy profile $\sigma$.

Rosenthal's potential function $\phi: \Sigma_1 \times \Sigma_2 \times \cdots \times \Sigma_n \rightarrow {\mathsf Z}$ is defined by
\begin{eqnarray}
\phi(\sigma) &=& \sum_{r\in{\cal R}} \sum_{i=1}^{n_r(\sigma)} g_r(i)  \label{eqn-potential} \\
&=& \sum_{i=1}^{N} \sum_{r\in\sigma_i} g_r(n_r^{i} (\sigma)) ~, \label{eqn-change-of-sums}
\end{eqnarray}
where the second equality comes from exchanging the two sums and $n_r^{i} (\sigma)$ denotes the number of players using resource $r$ whose index does not exceed $i$ (i.e., in the set $\{1, 2, \cdots, i\}$).

Now consider player $i$, who unilaterally moves from strategy $\sigma_i$ (corresponding to the profile $\sigma$) to strategy $\sigma_i^{'}$ (corresponding to the profile $\sigma^{'}$).  The potential changes by
\begin{eqnarray}
&&\Delta\phi (\sigma_i\rightarrow\sigma_i^{'}) \nonumber \\
&=& \sum_{r\in\sigma_i^{'}, r\not\in\sigma_i} g_r(n_r(\sigma)+1) -
\sum_{r\in\sigma_i, r\not\in\sigma_i^{i}} g_r(n_r(\sigma)) \nonumber \\
&=& \sum_{r\in\sigma_i^{'}} g_r(n_r(\sigma^{'})) - \sum_{r\in\sigma_i} g_r(n_r(\sigma)) \nonumber \\
&=& g^i(\sigma^{-i}, \sigma_i^{'}) - g^{i}(\sigma^{-i}, \sigma_i) ~,
\end{eqnarray}
where the second equality comes from the fact that for resources that are used by both strategies $\sigma_i$ and $\sigma_i^{'}$ there is no change in their total number of users.  The above result may be obtained either directly from Rosenthal's potential definition (\ref{eqn-potential}), or more easily, from the alternative change of sums equation (\ref{eqn-change-of-sums}) by assuming we are considering the $N$-th player.

The above result shows that the gain (loss) caused by any player's unilateral move is exactly the same as the gain (loss) in the potential, which may be viewed as a global objective function.  Since the potential of any strategy profile is finite, it follows that every sequence of improvement steps is finite, known as the finite improvement property (FIP), and they converge to a pure strategy Nash Equilibrium.  This NE is a local maximum (minimum) point of the potential function $\phi$, defined as a strategy profile where changing one coordinate cannot result in a greater value of $\phi$.

To summarize, we see that in this game, any sequence of unilateral improvement steps converges to a pure strategy NE, which is also a local optimum point of a global objective given by the potential function.

The $\phi()$ defined above is called an exact potential function, where individual payoff (cost) change as a result of a unilateral move is exactly reflected in this global function:
\begin{eqnarray}
g^i(\sigma^{-i}, \sigma_i^{'}) - g^{i}(\sigma^{-i}, \sigma_i) = \phi(\sigma^{-i}, \sigma_i^{'}) - \phi(\sigma^{-i}, \sigma_i)~.
\end{eqnarray}
More generally, a function $\phi$ is called an ordinal potential function if we have
$g^i(\sigma^{-i}, \sigma_i^{'}) \geq g^{i}(\sigma^{-i}, \sigma_i) \Leftrightarrow \phi(\sigma^{-i}, \sigma_i^{'}) \geq \phi(\sigma^{-i}, \sigma_i)$.  Games that possess the above properties are called exact potential games and ordinal potential games, respectively.
A congestion game is thus an exact potential game.  In \cite{Monderer1996} it was also shown that every potential game may be converted into an equivalent congestion game.  



\subsection{Extension to Resource Reuse}

It should now be clear why the standard definition of a congestion game does not capture the features of wireless communication.  In particular, if we consider channels as resources, then the payoff $g_r(n)$ for using channel $r$ when there are $n$ simultaneous users does not reflect reality: the function $g_r()$ is user specific in that the quantity $n$ is perceived differently by different users, depending on how many {\em interfering} users a user has.  This user specificity is also different from that studied in \cite{milchtaich1996cgp}, where $g_r()$ is a user-specific function $g_r^{i}$ but it takes the non-user specific argument $n$.  In other words, while the user-specific payoff is reflected in the functional form of the payoff function, in our context it is reflected through the user-specific argument. 
 
In our recent paper \cite{liu-allerton08} we took the approach of adopting alternative definitions of resources to circumvent some of the above problem.  In particular, by defining resources as certain spectrum-space units we were able to map some existing formulations on interference minimization into standard congestion game forms, and therefore were able to directly apply properties associated with congestion games.  However, a major limitation of this approach is that it does not work well when user objectives are rate maximizing rather than interference minimizing.  To understand what happens when the user objective is rate maximization, where a user's payoff $g_r(n)$ for using channel $r$ where there are a total of $n$ interfering users (including itself) is a non-increasing function of $n$, we would need to direct extend and generalize the definition of the standard congestion game.

For the rest of this paper, the term {\em player} or {\em user} specifically refers to a {\em pair} of transmitter and receiver in the network.  Interference in this context is between one user's transmitter and another user's receiver.  This is commonly done in the literature, see for instance \cite{Tse2007}.  We will also assume that each player has a fixed transmit power.  
%

\section{Problem Formulation}
\label{sec-problem}

In this section we formally definite our generalized congestion games, also referred to as {\em congestion games with resource reuse}.

Specifically, CG-RR has one more element than the standard CG.  It is given
$({\cal N}, {\cal R}, (\Sigma_i)_{i\in\cal{N}}, \{N_i\}_{i\in\cal{N}}, (g_r)_{r\in\cal{R}})$, where $N_i$ is the interference set of user $i$, including itself, while all other elements maintain the same meaning as before.  The payoff user $i$ receives for using resource $r$ is given by $g_r(n_r^i(\sigma))$ where $n_r^i(\sigma) = |\{j: r \in \sigma_j , j \in N_i \}|$.  That is, user $i$'s payoff for using $r$ is a function of the number of users interfering with itself, including itself.
%

A user's payoff is the summation of payoffs from all the channels he is using.
Note that if a user is allowed the strategy to simultaneous use all available resources, then its best strategy is to simply use all of them regardless of other users, provided that $g_r$ is a non-increasing function.  If all users are allowed such a strategy, then the existence of an NE is trivially true.

In this paper, we will limit our attention to the special case where each user is allowed only one channel at a time, i.e., its strategy space consists of $R$ single channel strategies.
In this case the payoff user $i$ receives for using a single channel $r$ is given by $g_r(n_r^i)$ where $n_r^i(\sigma)=|\{j: r = \sigma_j , j \in N_i \}|$.
Our goal is to find out what property this game has, in particular, when does an NE exist.
Other issues of interest to us include whether or not this game is a potential game, whether or not it has the finite improvement property. 
%
It's worth noting that due to this generalization, Rosenthal's definition of a potential function as given in the previous section no longer applies.

By the definition of the function $g_r(n)$ of resource $r$, it is implied that all users have the same payoff function when they use $r$ (they may perceive different values of $n$, but the function applies to all).  If this function is different to different users, i.e., given by $g^i_r(n)$, then we refer to this as the {\em user-specific} payoff functions.  In our motivating application this may be interpreted as users with different coding/modulation schemes may obtain different rates from using the same channel.  In subsequent sections some of our analysis is limited to the non-user-specific payoff function, while others can be generalized to the case of user-specific payoff functions.  This is shown through the difference in the notation $g_r(n)$ vs. $g^i_r(n)$.

To slightly simplify this problem, we make the extra assumption that $i \in N_j$ if and only if $j \in N_i$. This has the intuitive meaning that if one node $i$ interferes with another node $j$, the reverse is also true.  This symmetry does not always hold in reality, but is nonetheless a useful one to help obtain meaningful insight.  We explicitly assume that payoff function for any channel is non-increasing in the number of perceived interfering users. 


It is easy to see that we can equivalently represent the problem on a graph, where each node represents a user and there is a directed edge leaving node $i$ and entering node $j$ only if $i \in N_j$.  This can now be phrased as a coloring problem where each node needs to pick a color and receive a value depending on conflict (number of same colors neighboring to a node), and where the goal is to see whether a decentralized selfish scheme leads to an NE.
For the special case that we consider in this paper, the graph is undirected, where there is an edge between nodes $i$ and $j$ only if $i \in N_j$ and $j \in N_i$.

For simplicity of exposition, in subsequent sections we will often present the problem in its coloring version, and will use the terms {\em resource}, {\em channel}, and {\em color} interchangeably.

\section{Existence of the Finite Improvement Property}
\label{sec-fip}

In this section we investigate whether the CG-RR possesses the FIP property.  Once a game has the FIP, it immediately follows that it has an NE as we described in Section \ref{sec-review}.
Below we show that in the case of two resources (colors) the CG-RR game indeed has the FIP property, and as a result an NE exists.   Furthermore, this property holds in the case of two resources even when the payoff functions are user-specific.

We also show through a counter example that for the case of $3$ or more colors the FIP property does not hold.  This also implies that in such cases an exact potentially function does not exist for this game, as the FIP is a direct consequence of the existence of a potential function.  



\subsection{Finite Improvement Property for 2 resources}

In this section we prove that the finite improvement property holds when there are only two resources/colors to choose from and a user can only use one at a time. We shall establish this result by a contradiction argument. Suppose that we have a sequence of updates (we will remove the word {\em asynchronous} in the following with the understanding that whenever we refer to updates they are assumed to be asynchronous updates) that starts and ends in the exact same color assignment (or state) for any user. We denote such a sequence by
\begin{eqnarray}
U = \{u(1), u(2), \cdots u(T)\},
\end{eqnarray}
where $u(t)\in \{1, 2, \cdots, N\}$ denotes the user making the change at time $t$, and $T$ is the length of this sequence.
The starting state (or the color choice) of the system is given by
\begin{eqnarray}
S(1) = \{s_1(1), s_2(1), \cdots, s_N(1)\},
\end{eqnarray}
where $s_i(1) \in \{ r, b\}$, i.e., the state of each user is either ``r'' for Red, or 	``b'' for Blue.  We assume that a user's state/color is observed at time $t^{-}$, i.e., right before a color change is made by some user at time $t$.  In other words, $s_i(t)$ denotes the color of user $i$ at time $t^{-}$.  We use the notation $\bar{s}$ to denote the opposite color of a color $s$.

\begin{figure}
\centering
\includegraphics[width=2.0in]{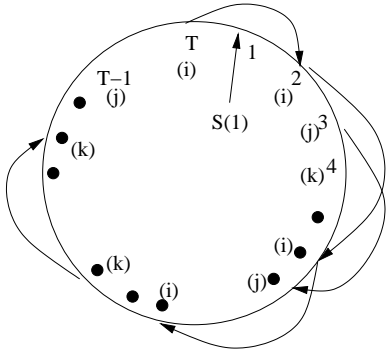}
\caption{Representing an improvement loop on a circle: times of updates $t$ and the updating user $u(t)$ are illustrated.}
\label{fig-circle}
\end{figure}

Since this sequence of updates form a loop in that $S(1^{-})=S(T^{+})$, we can naturally view these updates on a circle, starting at time $1^{-}$ and ending at $T^{+}$, when the system returns to its original state.  This is shown in Figure \ref{fig-circle}. Note that traversing the circle starting from any point gives rise to an improvement path; hence the notion of a starting point becomes inconsequential.

Since this sequence of updates is an improvement path, each change must not decrease the payoff of the user making the change. For example, suppose user $i$ changes from red to blue at time $t$, and $i$ has $x$ red neighbors and $y$ blue neighbors at $t$. Then we must have:
\begin{align}
g_b^{i}(y+1) > g_r^{i}(x+1),
\end{align}
where we denote user $i$'s specific payoff function as $g^i_r()$ and $g^i_b()$, respectively, for colors $r$ and $b$. Similarly, we can obtain one inequality for each of the $T$ changes. Our goal is to show that these $T$ inequalities cannot be consistent with each other.
The challenge here is that this contradiction has to hold for arbitrary non-increasing functions $\{g_r^{i},g_b^{i}\}$.  The way we address this challenge is to show that the above inequality leads to another inequality that does {\em not} involve the payoff functions when we consider pairs of reverse changes by the same user.  This is shown in Lemma~\ref{lemma:reversechange}. 


\begin{definition}[Reverse-change pairs]~\\
Consider an arbitrary user $i$'s two reverse strategy/color changes in an improvement path, one from $s$ to $\bar{s}$ at time $t$ and the other from $\bar{s}$ to $s$ at time $t'$. Let $\mathcal{SS}_{t,t'}$ denote the set of $i$'s neighbors not including $i$ who have the same color as $i$ at both times of change (i.e., at $t^{-}$ and $t'^{-}$, respectively).  Let $\mathcal{OO}_{t,t'}$ denote the set of $i$'s neighbors not including $i$ who have the opposite color as $i$ at both times of change. Similarly, we will denote by $\mathcal{SO}_{t,t'}$ (respectively $\mathcal{OS}_{t,t'}$) the number of $i$'s neighbors whose color is the same as (respective opposite of) $i$'s at the first update and the opposite of (respectively same as) $i$'s at the second update.
\label{def:SSOO}
\end{definition}

\begin{lemma} {\bf (Reverse-change inequality)}
Consider the CG-RR game with two resources/colors. Consider an arbitrary user $i$'s two reverse strategy/color changes in an improvement path, one from $s$ to $\bar{s}$ at time $t$ and the other from $\bar{s}$ to $s$ at time $t'$.  Then we have
\begin{align}
|\mathcal{SS}_{t,t'}| > |\mathcal{OO}_{t,t'}|,
\end{align}
That is,
among $i$'s neighbors not including $i$, there are strictly more users that have the same color as $i$ at both times of change, than those with the opposite color as $i$ at both times of change.
\end{lemma}
\label{lemma:reversechange}
\begin{proof} Since this is an improvement path, whenever $i$ makes a change it's for higher payoff.  Thus we must have at the time of its first change and its second change, respectively, the following inequalities:
\begin{align}
g^i_{\bar{s}}(|\mathcal{OS}_{t,t'}|+|\mathcal{OO}_{t,t'}|+1) &> g^i_{s}(|\mathcal{SO}_{t,t'}|+|\mathcal{SS}_{t,t'}|+1)\label{eqn-lemma1-1}\\
g^i_s(|\mathcal{SO}_{t,t'}|+|\mathcal{OO}_{t,t'}|+1) &> g^i_{\bar{s}}(|\mathcal{OS}_{t,t'}|+|\mathcal{SS}_{t,t'}|+1)\label{eqn-lemma1-2}
\end{align}
We now prove the lemma by contradiction.  Suppose that the statement is not true and that we have $|\mathcal{SS}_{t,t'}| \le |\mathcal{OO}_{t,t'}|$.  We then have
\begin{align}
g^i_{\bar{s}}(|\mathcal{OS}_{t,t'}|+|\mathcal{SS}_{t,t'}|+1) &\geq  g^i_{\bar{s}}(|\mathcal{OS}_{t,t'}|+|\mathcal{OO}_{t,t'}|+1) \nonumber\\
&> g^i_s(|\mathcal{SO}_{t,t'}|+|\mathcal{SS}_{t,t'}|+1) \nonumber\\
&\geq g^i_s(|\mathcal{SO}_{t,t'}|+|\mathcal{OO}_{t,t'}|+ 1)
\end{align}
where the first and the third inequalities are due to the non-increasing assumption on the payoff functions, and the second inequality is due to (\ref{eqn-lemma1-1}).  However, this contradicts with (\ref{eqn-lemma1-2}), completing the proof.
\end{proof}

We point out that by the above lemma the payoff comparison is reduced to counting different sets of users.  This greatly simplifies the process of proving the main theorem of this section.  Below we show that it is impossible to have a finite sequence of asynchronous improvement steps ending in the same color state (set of user strategies) as it started with.  At the heart of the proof is the repeated use of the above lemma to show that loops cannot form in a sequence of asynchronous updates.

\begin{theorem}\label{thm:fip-2color}
When there are only two resources/colors to choose from and a user can only use one at a time, we have the finite improvement property.
\end{theorem}
\begin{proof}
We prove this by contradiction.  As illustrated by Figure~\ref{fig-circle}, we consider a sequence of improvement updates that results in the same state. 

Consider every two successive color changes, along this circle clockwise starting from time $t=1$, that a user $u(t)$ makes at time $t$ and $t'$ from color $s=s_{u(t)}(t)$ to $\bar{s}$, and then back to $s$, respectively.  Note that this will include the two ``successive'' changes formed by a user's last change and its first change (successive on this circle but not in terms of time).  We have illustrated this in Figure \ref{fig-circle} by connecting a pair of successive color changes using an arrow. It is easy to see that there are altogether $T$ arrows.

For each arrow in Figure \ref{fig-circle}, or equivalently each pair of successive color changes by the same user, we consider the two sets $\mathcal{SS}_{t,t'}$ and $\mathcal{OO}_{t,t'}$ in Definition~\ref{def:SSOO}. Due to the user association, we will also refer to these sets as {\em perceived} by user $u(t)$.  
By Lemma 1, given an updating sequence with the same starting and ending states, we have for each pair of successive reverse changes by the same user, at time $t$ and time $t'$, respectively:
\begin{eqnarray}\label{eqn-inequalities}
|\mathcal{SS}_{t,t'}| > |\mathcal{OO}_{t,t'}|. 
\end{eqnarray}
That is, $\mathcal{SS}$ sets are strictly larger than $\mathcal{OO}$ sets.  

This gives a total of $T$ inequalities, one for each update in the sequence and each containing two sets.  Equivalently there is one inequality per arrow illustrated in Figure \ref{fig-circle}.
We next consider how many users are in each of these $2T$ sets (note that by keeping the same ``$>$'' relationship, the $\mathcal{SS}$ sets are always on the LHS of these inequalities and the $\mathcal{OO}$ sets are always on the RHS).  To do this, we will examine users by pairs -- we will take a pair of users and see how many times they appear in each other's sets in these inequalities.  In Claim~\ref{claim} below, we show that they collectively appear the same number of times in the LHS sets and in the RHS sets.  We then enumerate all user pairs.  What this result says is that these users collectively contributed to an equal number of times to the LHS and RHS of the set of inequalities given in Eqn (\ref{eqn-inequalities}). Adding up all these inequalities, this translates to the fact that the total size of the sets on the LHS and those on the RHS much be equal.  This however contradicts the strict inequality, thus completing the proof. 
\end{proof}

\begin{claim}
Consider a pair of users $A$ and $B$ in an improvement updating loop, and consider how they are perceived in each other's set.  Then A and B collectively appear the same number of times in the LHS sets (the $\mathcal{SS}$ sets) and in the RHS sets (the $\mathcal{OO}$ sets).
\label{claim}
\end{claim}
\begin{proof}
First note that $A$ and $B$ have to be in each other's interference set for them to appear in each other's $\mathcal{SS}$ and $\mathcal{OO}$ sets.  Since we are only looking at two users and how they appear in each other's sets, without loss of generality we can limit our attention to a subsequence of the original updating sequence involving only $A$ and $B$, given by
\begin{eqnarray}
U_{AB} = \{ u(t_1), u(t_2), \cdots, u(t_l)\}
\end{eqnarray}
where $u(t_i) \in \{A, B\}$, $t_i \in\{1, 2, \cdots, T\}$, and $l$ is the length of this subsequence, i.e., the total number of updates between $A$ and $B$.  As before, this subsequence can also be represented clockwise along a circle.

It helps to consider an example of such a sequence, say, ABAABBABAA, also shown in Figure \ref{fig-example}.  In what follows we will express an odd train as the odd number of consecutive changes of one user sandwiched between the other user's changes, e.g., the odd train ``BBB'' in the subsequence ``ABBBA''.
\begin{figure}
\centering
\includegraphics[width=2.0in]{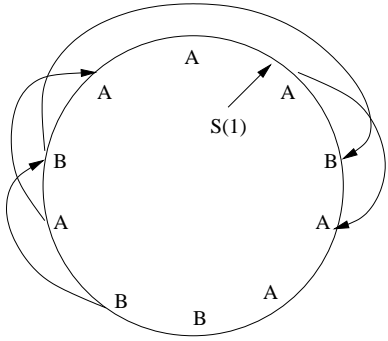}
\caption{Example of an updating sequence ``ABAABBABAA'' illustrated on a circle.}
\label{fig-example}
\end{figure}

A few things to note about such a sequence:
\begin{enumerate}
\item Since the starting and ending states are the same, each user must appear an even number of times in the update sequence.
Since each user appears an even number of times, there must be an even number of odd trains along the circle for any user. 

\item A user (say A) only appears in the other's (say B's) $\mathcal{SS}$ or $\mathcal{OO}$ sets if it has an odd train between the other user's two successive appearances.   This means that there is an even number of relevant inequalities where A appears in B's inequalities (either on the LHS or the RHS), and vice versa.

\item Consider the collection of all relevant inequalities discussed above one for each odd train, in the order of their appearance on the circle (all four such inequalities are illustrated in Figure \ref{fig-example}).  Then A and B contribute to each other's inequalities on alternating sides along this updating sequence/circle.  That is, suppose the first inequality is A's and B goes into its LHS, then in the next inequality (could be either A's or B's) the contribution (either A to B's inequality or B to A's inequality) is on the RHS.  Take our running example, for instance, the first inequality is due to the odd train marked by the sequence ABA, and the second BAB.  Suppose A and B start with different colors, then in the first inequality, B appears in the RHS; in the second, A appears in the LHS.
\end{enumerate}

We now explain why the third point above is true.
The reason is because for one user (B) to appear in the other's (A's) LHS, they must start by having the same color and again have the same color right before A's second change (see e.g. the subsequence ``ABA'' in the running example).  Until the next odd train (``BAB''), both will make an even number of changes including A's second change (``AABB''). The next inequality belongs to the user who makes the last change before the odd train (B) .  As perceived by this user (B) right before this change, the two must now have different colors.  This is because as just stated A will have made an even number of changes from the last time they are of the same color (by the end of ``AB''), while B is exactly one change away from an even number of changes (by the end of ``ABAAB'').  Therefore, the contribution from the other user (A) to this inequality must be to the RHS.

Alternatively, one can see that essentially the color relationship between A and B reverses upon each update, and there is an odd number of updates between the starting points of two consecutive odd trains, so the color relationship flips for each inequality in sequence.

The above argument establishes that as we go down the list of inequalities and count the size of the sets on the LHS vs. that on the RHS, we alternate between the two sides.  Since there are exactly even number of such inequalities, we have established that A and B collectively appear the same number of times in the LHS sets and in the RHS sets.
\end{proof}

\subsection{Counter-Example for 3 Resources}
The above theorem establishes that when there are only two resources, the FIP property holds, and consequently an NE exists.  This holds for the general case of user-specific payoff functions.  Below we show a counter-example that the FIP property does not necessarily hold for $3$ resources/colors or more.

\begin{figure}[htb!]
\centering%
\includegraphics[width=7cm]{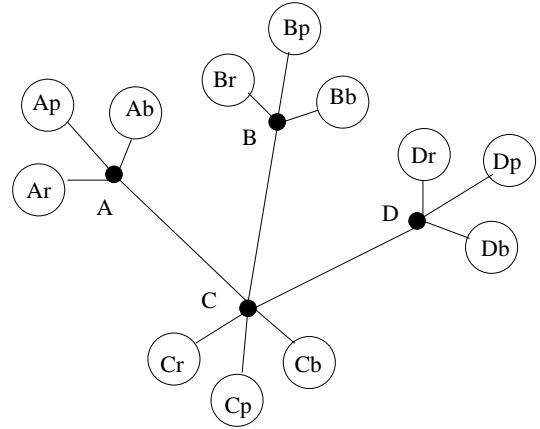}
\caption{A counter example of 3 colors: nodes $A$, $B$, $C$, and $D$ are connected as shown; in addition, node $W$,  $W\in\{A, B, C, D\}$, is connected to $W_x$ other nodes of color $x\in\{r, p, b\}$. }
\label{fig:counter}
\end{figure}

\begin{example}
Suppose we have three colors to assign, denoted by $r$  (red), $p$ (purple), and $b$ (blue).  Consider a network topology shown in Figure \ref{fig:counter}, where we will primarily focus on nodes $A$, $B$, $C$ and $D$.
In addition to node $C$, node $A$ is also connected to $A_r$, $A_p$ and $A_b$ nodes of colors red, green and blue, respectively.   $B_r$, $B_p$, $B_b$, $C_r$, $C_p$, $C_b$, and $D_r$, $D_p$, $D_b$ and similarly defined and illustrated in Figure \ref{fig:counter}.  Note that there may be overlap between these quantities, e.g., a single node may contribute to both $A_r$ and $B_r$, and so on.

Consider now the following sequence of improvement updates involving only nodes $A$, $B$, $C$, and $D$, i.e., within this sequence none of the other nodes change color (note that this is possible in an asynchronous improvement path), where the notation $s_1\rightarrow s_2$ denotes a color change from $s_1$ to $s_2$ and at time $0$ the initial color assignment is given. 

\medskip
\begin{center}
\begin{tabular}{|r|c|c|c|c|}
\hline
time step & $A$ & $B$ & $C$ & $D$ \\
\hline
0 & b 			 & p & p & b \\
\hline
1 & b $\rightarrow$ r &    &     &  \\
\hline
2 &                            & p $\rightarrow$ r & & \\
\hline
3 &                            &                            &                   & b $\rightarrow$ r \\
\hline
4 & 				& 				& p $\rightarrow$ r &                 \\
\hline
5 & r $\rightarrow$ p &  & & \\
\hline
6 & 				& 				& 				& r $\rightarrow$ b \\
\hline
7 & 				& r $\rightarrow$ b & & \\
\hline
8 & 				& 				& r $\rightarrow$ b & \\
\hline
9 & p $\rightarrow$ b & & & \\
\hline
10& 				& 				& b $\rightarrow$ p & \\
\hline
11& 				& b $\rightarrow$ p & & \\
\hline
\end{tabular}
\end{center}

\medskip 
\medskip 
We see that this sequence of color changes form a loop, i.e., all nodes return to the same color they had when the loop started.
%
%
%
For this to be an improvement loop such that each color change results in improved payoff, it suffices for the following sets of conditions to hold (here we assume all users have the same payoff function):
\begin{eqnarray*}
&& g_r(A_r+1) > g_b(A_b+1) > g_b(A_b+2) \\
&>& g_p(A_p+1) > g_r(A_r+2) ~; \\
&& g_r(B_r+1) > g_p(B_p+2) > g_b(B_b+1) \\
&>& g_r(B_r+2) ~; \\
&& g_b(C_b+3) > g_r(C_r+1) > g_r(C_r+4) \\
&>& g_p(C_p+1)>g_b(C_b+4) ~; \\
&& g_r(D_r+1) > g_b(D_b+1) > g_r(D_r+2)
\end{eqnarray*}
It is straightforward to verify the sufficiency of these conditions by following a node's sequence of changes.

To complete this counter example, it remains to show that the above set of inequalities are feasible given appropriate choices of $A_x$, $B_x$, $C_x$ and $D_x$, $x\in\{r, p, b\}$.  There are many such choices, below we give one example:
\begin{eqnarray*}
A_x = 5; ~~~ B_x = 3; ~~~ C_x = 7; ~~~ D_x = 1, ~~~ x \in\{ r, p, b\}
\end{eqnarray*}
With such a choice, and substituting them into the earlier set of inequalities, we obtain the following single chain of inequalities:
\begin{eqnarray*}
&& g_r(2) > g_b(2) > g_r(3) \\
&>& g_r(4) > g_p(5) > g_b(4) > g_r(5) \\
&>& g_r(6) > g_b(6) > g_b(7) > g_p(6) > g_r(7) \\
&>& g_b(10) > g_r(8) > g_r(11) > g_p(8)>g_b(11)
\end{eqnarray*}
It should be obvious that this chain of inequalities can be easily satisfied by the right choices of payoff functions.
%
\end{example}

It is easy to see how if we have more than 3 colors , this loop will still be an improving loop as long as the above inequalities hold. This means that for 3 colors or more the FIP property does not hold.  Note that the updates in this example are not always best response updates .
It is also now obvious from this example that potential and semi-potential functions don't always exist for cases with 3 colors or more.

\section{Sufficient Conditions on Graph Properties}
\label{sec-graph}

In this section we examine what graph properties will guarantee the existence of an NE. 
The results here are not as general as we would've liked, though they do shed further insights on the nature of this class of games. 

\begin{theorem}
When the graph is complete a NE always exists for the CG-RR game defined on this graph. 
\end{theorem}

This theorem has a trivial proof.  It is simply a direct consequence of known results on the standard CG: in a complete graph every node is every other node's neighbor, therefore the CG-RR extension reduces to the original CG, thus this result.  Furthermore, for the same reason when the graph is complete the FIP property also holds.

\begin{theorem}
Consider the network graph in the form of a tree, and order the $R$ resources/colors in decreasing order of $g_r(1)$, such that $g_1(1)\geq g_2(1)\geq \cdots \geq g_R(1)$.  If $g_2(1) \geq g_1(2)$, then a Nash equilibrium exists.
\end{theorem}

\begin{proof}
Consider an arbitrary node $A$ as the root of the tree, and partition all other nodes according to their distance to node $A$ into sets $A_1, A_2, \cdots, A_L$, where set $A_i$ contains all nodes that at exactly distance $i$ away from node $A$, and $L$ is the depth of the tree.  We will refer to sets $A_i$ where $i$ is even as {\em even sets}, and refer to sets $A_i$ where $i$ is odd as {\em odd sets}. 
Now assign resource $1$ to all nodes in sets $A_i$ where $i$ is even; assign resource $2$ to all nodes in sets $A_i$ where $i$ is odd.  We claim this is a Nash equilibrium assignment. 

Note that due to the tree structure, any node in an even set has all its neighbors in odd sets, and vice versa.  As a result, a node that belong to an even set has a payoff $g_1(1)$ and has no incentive to change its strategy.  A node that belong to an odd set has a payoff $g_2(1)$ and is connected to at least one node with resource $1$.   Under the condition that $g_2(1)>g_1(2)$, this node also has no incentive to change its strategy, thus completing the proof. 

\end{proof}

\begin{theorem}
When the network has a star topology where a single node $A$ is connected to all other nodes $A_1,A_2,\ldots,A_{N-1}$, a Nash equilibrium exists. 
\end{theorem}

\begin{proof}
As in the previous theorem, we order the resources/colors such that $g_1(1)\geq g_2(1) \geq \cdots \geq g_R(1)$.  

If $g_2(1) \geq g_1(N)$ then we assign $2$ to node $A$ and assign $1$ to all other nodes. This is trivially an NE.

If $g_1(N) \geq g_2(1)$ then we assign $1$ to all nodes and this again is an NE. 
\end{proof}


\begin{theorem}
If the network is in the form of a loop, then there always exists a Nash equilibrium involving no more than 3 resources/colors.
\end{theorem}

\begin{proof}
We know from Theorem \ref{thm:fip-2color} that when there are only 2 colors an NE always exists.  Thus assume there are at least 3 colors to choose from. 
Thus there always exist three colors $r, b, p$ that have the highest single-user occupancy payoff values and suppose we have
\begin{eqnarray*}
g_r(1) \geq g_b(1) \geq g_p(1)
\end{eqnarray*}

If the loop has an even number of nodes then compare $g_r(3)$ with $g_b(1)$.  If $g_r(3)\geq g_b(1)$, then assigning $r$ to all nodes will result in an NE; if $g_b(1)\geq g_r(3)$ then assigning $r$ and $b$ alternately will result in an NE. 

Now consider the case where the loop has an odd number of nodes, labeled from $1$ to $2n+1$, where node $i$ is connected to node $i+1$ and node $2n+1$ is connected to node $1$.  Again we see that if $g_r(3)\geq g_b(1)$ then assigning $r$ to all nodes results in an NE. 

Assume now $g_b(1)\geq g_r(3)$ and consider the following assignment.  Assign $r$ and $b$ alternately to nodes from $2$ up to $2n$, so that nodes $2$ and $2n$ are both colored $r$.  It remains to determine the coloring of nodes $1$ and $2n+1$.  
We have the following four cases (under the condition $g_b(1)\geq g_r(3)$):

\begin{enumerate}

\item $g_b(2)\geq g_r(2)$ and $g_b(2)\geq g_p(1)$: in this case ($b$, $b$) assignment to nodes $1$ and $2n+1$ will result in an overall NE. 

\item $g_b(2)\geq g_r(2)$ and $g_b(2) < g_p(1)$: in this case either ($b$, $p$) or ($p$, $b$) for nodes $1$ and $2n+1$ will result in an overall NE. 

\item $g_b(2) < g_r(2)$ and $g_r(2)\geq g_p(1)$: in this case either ($b$, $r$) or ($r$, $b$) for node $1$ and $2n+1$ will result in an overall NE. 

\item $g_b(2) < g_r(2)$ and $g_r(2) < g_p(1)$: in this case either ($b$, $p$) or ($p$, $b$) for nodes $1$ and $2n+1$ will result in an overall NE. 
\end{enumerate}

Therefore in all cases we have shown an NE exists. 

\end{proof}  

\begin{corollary}
In a chain network (an open loop), an NE exists that involve no more than 2 colors. 
\end{corollary}


\section{Sufficient Conditions on User Payoff Functions}
\label{sec-payoff}

In this section we examine what properties on the user payoff functions will guarantee the existence of an NE.  Specifically, we show that for general network graphs, an NE always exists if (1) there is one resource with a dominating payoff function (much larger than the others), or (2) different resources present the same type of payoff for users.  Moreover, in the case of (2) the game has the FIP property.  We note that case (2) is of particular practical interest and relevance, as this case in the context of spectrum sharing translates to evenly dividing a spectrum band into sub-bands, each providing users with the same bandwidth and data rate.  Below we present and prove these results. 

\begin{theorem}
For a general network graph, if there exists a resource $r$ and its payoff function is such that $g_r(N_{d}) \geq g_s(1)$, where $N_d=\max\{N_i, i=1, 2, \cdots, N\}$, for all $s \in {1,2,\cdots,R}$ then a Nash Equilibrium exists.
\end{theorem}

Here $N_d$ is the maximum node degree in the network, i.e., the maximum possible number of users sharing the same resource.  In words, this theorem says that if there exists a resource whose payoff ``dominates'' all other resources, an NE exists.  This is a rather trivial result; an obvious NE is when all users share the dominating resource. 

\begin{theorem}
For a general network graph, if all resources have identical payoff functions, i.e., for all resources $r$ and $s$, we have $g_r(n)=g_s(n)=g(n)$ for $n=1, 2, \cdots, N$ and some function $g(\dot)$, then there exists a Nash Equilibrium, and the game has the finite improvement property.
\end{theorem}

\begin{proof}
We prove this theorem by using a potential function argument.  

Recall that user $i$'s total payoff under the strategy profile $\sigma$ is given by (here we have suppressed the subscript $r$ since all resources are identical): 
\begin{eqnarray}
g^{i}(\sigma) = g(n^{i}(\sigma)), ~~ n^{i}(\sigma) = |\{j: \sigma_j = \sigma_i, j\in N_i\}| ~ 
\end{eqnarray}
where $\sigma_i \in {\cal R}$ since we have limited our attention to the case where each user can select only one resource at a time. 

Now consider the following function defined on the strategy profile space: 
\begin{eqnarray}
\phi(\sigma) &=& \sum_{i, j \in{\cal N}} 1(i\in N_j) 1(\sigma_i=\sigma_j) \nonumber \\
&=& \frac{1}{2} \sum_{i\in {\cal N}} n^{i}(\sigma) ~, 
\end{eqnarray}
where 
the indicator function $1(A)=1$ if $A$ is true and $0$ other wise.  For a particular strategy profile $\sigma$ this function $\phi$ is the sum of all pairs of users that are connected (neighbors of each other) and have chosen the same resource under this strategy profile.  Viewed in a graph, this function is the total number of edges connecting nodes with the same color.  

We see that every time user $i$ improves its payoff by switching from strategy $\sigma_i$ to $\sigma_i^{'}$, and reducing $n^{i}(\sigma^{-i}, \sigma_i)$ to $n^{i}(\sigma^{-i}, \sigma_i^{'})$ ($g$ is a non-increasing function), the value of $\phi()$ strictly decreases accordingly\footnote{It's easily seen that a non-increasing function $G(\sum_{i, j \in{\cal N}} 1(i\in N_j) 1(\sigma_i=\sigma_j))$ is an ordinal potential function of this game as its value improves each time a user's individual payoff is improved thereby decreasing the value of its argument.}. 
%
%
As this function is bounded from below, this means that in this case the game has the FIP property so this process eventually converges to a fixed point which is a Nash Equilibrium.  
 \end{proof}

\section{Conclusion}
\label{sec-conclusion}

In this paper we have considered an extension to the classical definition of congestion games by allowing resources to be reused among non-interfering users.  This is a much more appropriate model to use in the context of wireless networks and spectrum sharing where due to decay of wireless signals over a distance, spatial reuse is frequently exploited to increase spectrum utilization. 

The resulting game, congestion game with resource reuse, is a generalization to the original congestion game.  We have shown that when there are only two resources and users can only use one at a time, then the game has the finite improvement property; the same is shown to be false in general when there are three or more resources.  We further showed a number of conditions on the network graph as well as the user payoff functions under which the game has an NE.  Perhaps most relevant to spectrum sharing is the result that when all resources present the same payoff to users (e.g., all channels are of the same bandwidth and data rate for all users), then the game has the finite improvement property and an NE exists.

\section{ACKNOWLEDGMENTS}

The authors gratefully acknowledge helpful discussions with Prof. Jianwei Huang.


\end{document}